\documentclass[10pt]{article}
\usepackage{geometry}                
\usepackage{graphicx}
\usepackage{color}
\usepackage{amsmath}
\usepackage{amssymb}
\usepackage{amsbsy}
\newcommand{\homog}{$\pm \varphi_0$}
\newcommand{\Sone}{{\bf s_{1}}}
\newcommand{\Stwo}{{\bf s_{2}}}
\newcommand{\Sthree}{{\bf s_{3}}}
\newcommand{\rparen}{({\bf r})}

\newcommand{\kparen}{({\bf k})}

\newcommand{\pparen}{({\bf p})}

\newcommand{\sumn}{\sum\limits}

\newcommand{\bfr}{{\bf r}}

\newcommand{\bfk}{{\bf k}}

\newcommand{\bfp}{{\bf p}}
\newcommand{\bfKpm}{{\bf K_{\pm}}}

\newcommand{\apm}{{a_{\pm}}}
\newcommand{\bpm}{{b_{\pm}}}
\newcommand{\sigvec}{\boldsymbol \sigma}
\newcommand{\gammavec}{{\boldsymbol \gamma}}

\newcommand{\alphavec}{{\boldsymbol \alpha}}
\newcommand{\Avec}{{\bf A}}

\newcommand{\Dvec}{{\bf D}}
\title{Chiral Gauge Theory for Graphene}
\author{R. Jackiw$^*$ and S.-Y. Pi$^\dagger$\\
\it\small $^*$Department of Physics\\[-.5ex]
\it\small Massachusetts Institute of Technology\\[-.5ex]
\it\small Cambridge, MA 02139\\[1ex]
\it\small $^\dagger$ Department of Physics\\[-.5ex]
\it\small Boston University\\[-.5ex]
\it\small Boston, MA 02215\\
\small\tt MIT-CTP/3808 \hspace{1in} BUHEP-07-02}
\date{January 2007}                                           

\begin{document}

\maketitle

\begin{abstract}
We construct a chiral gauge theory to describe fractionalization of fermions in graphene. Thereby we extend a recently proposed model, which relies on vortex formation. Our chiral gauge fields provide dynamics for the vortices and also couple to the fermions.
\end{abstract}
\section{Introduction}
In some condensed matter systems the electron excitation spectrum near the Fermi surface can be described by a Dirac-type matrix equation. This equation does not arise from relativistic considerations, but rather by linearizing the energy dispersion near (a finite number of) Dirac points (intersections of the energy dispersion with the Fermi level). Such systems can exhibit fermion fractionalization if the Dirac equation possesses isolated bound states in the gap between negative-energy (valence band) states and positive-energy (conduction band) states.

A familiar example is 1-dimensional polyacetylene \cite{Jackiw:1975fn}. As is generally the case in one dimension, there are two Dirac points for polyacetylene. Therefore a 2-component Dirac equation governs electron motion near each Dirac point. A distortion of the underlying lattice (Peierls' instability) perturbes the electron motion in a way that couples the two Dirac points and opens a gap in the energy spectrum. In the Dirac equation description this is achieved by coupling the Dirac field $\Psi$ to a scalar field $\varphi$, which is a measure of the lattice distortion. $\varphi$ enjoys a $Z_2$ symmetry with two ground states in which it takes homogenous values $\pm \, \varphi_0$. This coupling to $\varphi_0$ leads to a Dirac mass $\propto | \varphi_0 |$. But $\varphi$ can also take  a position-dependent kink profile (soliton) $\varphi_s$ that interpolates between the two vacua $\pm \varphi_0$. This ``twisting" of the mass parameter describes a defect in the lattice distortion. The Dirac equation with the kink profile $\varphi_s$ replacing the homogenous \homog\, possesses a single mid-gap (zero-energy) eigenstate. This gives rise to fractional fermion number for the electrons: 1/2 per spin degree of freedom \cite{Jackiw:1975fn}.

Recently a similar story has been told by C.-Y. Hou, C. Chamon and C. Mudry [2] (HCM) about (monolayer) graphene. This is  a 2-dimensional honeycomb array of carbon atoms forming a hexagonal lattice, which may be viewed as a superposition of two triangular sublattices, A and B.
\begin{figure}[ht]
\begin{center}
\includegraphics[scale=.50]{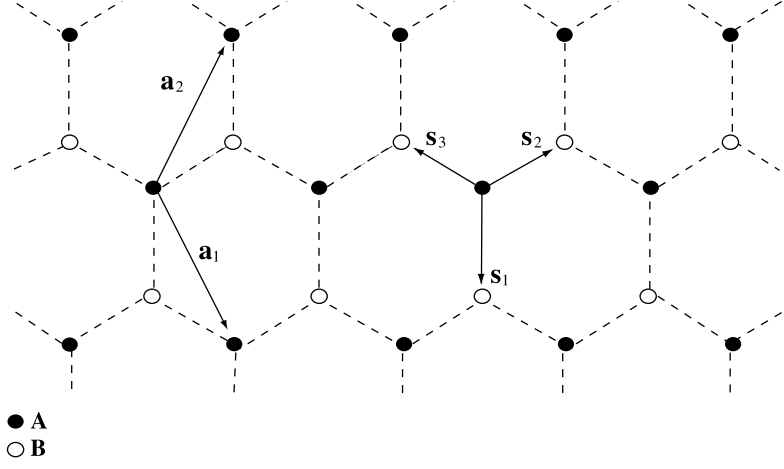}
\caption{Graphene hexagonal lattice constructed as a superposition of two triangular lattices {\bf A} and {\bf B}, with basis vectors $\bf a_{i}$ for lattice  {\bf A}  and vectors $\bf s_{i}$ connecting {\bf A} to {\bf B}.}
\label{sublattice}
\end{center}
\end{figure}
The generators of lattice A are ${\bf a_1} \ \mbox{and}\ {\bf a_2}$. The three vectors $\bf s_{i}$ connect any site from lattice A to its nearest neighbor sites belonging to B.
They are
\begin{equation}
\Sone = (0, -1) \ell, \quad\Stwo = \left({\frac{\sqrt{3}}{2}}, \frac{1}{2}\right) \ell,\quad \Sthree = \left(-{\frac{\sqrt{3}}{2}}, \frac{1}{2}\right) \ell
\label{rjeq1}
\end{equation}
where $\ell$ is the lattice spacing; see Figure 1.

When no lattice distortion is considered, the tight-binding Hamiltonian, with uniform hopping constant $t$, is taken as
\begin{equation}
H_{0} = - t\, \sumn_{{\bf r} \epsilon A} 
\ \sumn_{i = 1,2,3} \ \bigg(a^{\dagger} \, \rparen\, b ({\bf r} + {\bf s}_{i}) + b^{\dagger} ({\bf r} + {\bf s}_{i}) a \rparen  \bigg)\, ,
\label{rjeq2}
\end{equation}
where the fermion operators $a$ and $b$ act on sublattices A and B. $H_{0}$ is diagonal in momentum space.
\begin{equation}
\left\{\!\!\!\!\begin{array}{c}
a\, \kparen\\
b\, \kparen 
\end{array}\!\!\!\!
\right\} = \sumn_{{\bf r} \epsilon A} \ e^{-i \bfk \cdot \bfr} \  \left\{\!\!\!\!\begin{array}{c}
a\, \rparen\\
b\, \rparen 
\end{array}\!\!\!\!
\right\}
\label{rjeq3}
\end{equation}

\begin{subequations}
\begin{eqnarray}
H_{0} = \sum_{\bf k} \ \bigg(\Phi\, \kparen\, a^{\dagger} \, \kparen \, b\, \kparen + \Phi^{\ast} \, \kparen\, b^{\dagger} \, \kparen \, a\, \kparen\bigg)\label{rjeq4a}\\[1ex]
\Phi \, \kparen = - t\, \sumn_{i=1,2,3} e^{i \bfk \cdot {\bf s}_i }\hspace{.5in} \label{rjeq4b}
\end{eqnarray}
\end{subequations}
The single particle energy spectrum $E\kparen = \pm \mid \Phi \kparen \mid$ contains two zero-energy Dirac points at
\[
\bfk = \bfKpm = \pm \left(\frac{4\pi}{3\sqrt{3}\, \ell}, 0\right) ; \quad \Phi \, (\bfKpm) = 0.
\]
This parallels the 1-dimensional case, but is rare in two dimensions \cite{pw1947}.

$H_0$ is linearized around the two Dirac points, $\bfk = \bfKpm + \bfp$, and is supplemented by a term arising from a (Kekul\'{e}) distortion of the lattice, which couples the two Fermi points.
\begin{alignat}{4}
H&= \sum\limits_\bfp \ \bigg(\phi_+\, \pparen\, a^\dagger_+\pparen\, b_+\, \pparen + \phi^\ast_+\, \pparen\, b^\dagger_+\, \pparen\, a_+\, \pparen 
 + \phi_- \, \pparen\, a^\dagger_-\, \pparen \, b_-\, \pparen + \phi ^\ast\, \pparen\, b^\dagger_-\, \pparen \, a_-\, \pparen\bigg)\nonumber\\
& +\sum\limits_\bfp \ \bigg(\triangle_0 a^\dagger_+\, \pparen \, b_-\, \pparen + \triangle^\ast_0\, b^\dagger_-\, \pparen\, a_+\, \pparen
+ \triangle^\ast_0 \, a^\dagger_-\,  \pparen \, b_+ \, \pparen + \triangle_0\, b^\dagger_+ \, \pparen\, a_- \pparen \bigg)
\end{alignat}
$\phi_\pm$ is the linearization of $\Phi: \phi_\pm \pparen = \pm v_F\, (p_x \pm i \, p_y), v_F = 3 t \ell/2$, (hence forth $v_F$ is set to unity) and $a_\pm , b_\pm$ are fermion operators near the Dirac points: $\apm \pparen \equiv a\, (\bfKpm + \bfp), b_\pm ({\bf p})\equiv b (\bfKpm + \bfp)$. In the second sum $\triangle_0$ is a homogenous, but complex order parameter, which effects the coupling between the Fermi points $\bfKpm$ and leads to a mass gap in the single-particle energy dispersion:
$\epsilon\, \pparen = \pm \, \sqrt{\bfp^2 +\mid\triangle_0\mid^2}$ .

To find zero modes for this system HCM promote the mixing parameter $\triangle_0$ to an inhomogenous complex function $\triangle$ with a vortex profile.  To this end the Hamiltonian (5) is presented in coordinate space as
\begin{equation}
H =\int d^2 r \Psi^\ast\, \rparen\, K \Psi \rparen ,
\label{rjeq6}
\end{equation}
where $\Psi\, \rparen$ is a 4-component ``spinor"
\[
\Psi = \left(
\begin{array}{c}
\psi^b_+\\[1ex]
\psi^a_+\\[1ex]
\psi^a_-\\[1ex]
\psi^b_-
\end{array}
\right) \quad \mbox{with} \quad \vtop{\hsize= 1in $\psi^a_\pm \, \rparen = \int d^2\, p \, e^{-i \bfp \cdot \bfr}\, \apm\,\pparen$\\[1ex]
 $\psi^b_\pm\, \rparen = \int d^2 \, p \, e^{-i \bfp \cdot \bfr}\, \bpm\,\pparen$}
\]
and $K$ is the 4x4 matrix
\begin{equation}
K = \left(
\begin{array}{cccc}
0 & -2 i \partial_{z}  & \triangle \rparen & 0\\[1ex]
-2 i \partial_{z} \ast & 0 & 0 & \triangle \rparen \\[1ex]
\triangle^{\ast} \rparen & 0 & 0 & 2 i \partial_{z}\\[1ex]
0 & \triangle^{\ast}\rparen &2 i \partial_z\ast  &  0
\end{array}
\right)
\label{rjeq7}
\end{equation}
with $-2 i \partial_{z} = \frac{1}{i}\ (\partial_{x} -i \, \partial_{y})$.

HCM take $\triangle \rparen$ in the $n$-vortex form: $\triangle  (r)\, e^{i n \theta}$ where n is an integer, $\triangle (r)$ vanishes as $r ^{|n|}$ for small $r$, and approaches the mass-generating value $\triangle_0$ at large $r$. They then establish the occurrence of $|n|$ zero modes, {\it i.e.} solutions to $K \Psi = 0$, on lattice A (B) for negative (positive) $n$, and they construct explicitly the solution for $n= -1$. Therefore fermion number $\propto \int d^2 r \Psi^\ast \Psi$ is fractionalized.

\section{Chiral Gauge Theory for Graphene}
In this paper we elaborate the HCM model, and address the following two topics. HCM leave unspecified the dynamics that gives rise to the complex vortex profile. To remedy this, we  first   propose identifying the HCM vortex with the Nielsen, Olesen/Landau, Ginsburg, Abrikosov (NO/LGA) vortex, which is  described by a charged scalar field, as in the HCM model. However an $U (1)$ gauge field is also involved in creating a finite-energy NO/LGA vortex, but no gauge field is present in the HCM model. Therefore, second, we propose introducing a relevant gauge potential and coupling it to the Dirac fermions in a chiral manner.  This expands the symmetry of the interaction to the kinetic term, and renders the theory  invariant against local, chiral $U(1)$ gauge transformations, which also act on the scalar and Dirac fields. Additionally there remains a global fermion number $U(1)$ symmetry, and its charge is fractionalized.

To present our extension of the HCM model, we shall use Dirac matrix notation. We begin by rewriting $K$ in (7) in terms of Dirac matrices whose forms we take as follows.

\[
\alphavec = (\alpha^1, \alpha^2, \alpha^3) = {\sigvec\quad 0 \choose 0\, -\sigvec} \qquad \beta = {0\quad I \choose I\quad 0} 
\]
Note: we use 4x4 Dirac matrices, even though the minimal Dirac algebra in (2+1)-dimensions requires only 2x2 matrices. But we have four degrees of freedom: two each in lattices A and B. Since there are two spatial dimensions, we use only the first two $\alpha$ matrices : $\alpha^i , i = 1,2$ or $ x, y$; while the role of the third one, $\alpha^3$, will emerge later. Also unlike with minimal 2x2 Dirac matrices, here we can construct the chiral $\gamma_5$ matrix, as the Hermitian quantity
\[
\gamma_5 = -i \alpha^1\, \alpha^2\, \alpha^3 = {I\quad 0 \choose 0\, -I}, \quad \gamma^2_5 = I .
\]
The 
$\gamma$ matrices read
\[
\gammavec = \beta\, \alphavec = {0\,-\sigvec  \choose \sigvec\quad 0} ,\quad \gamma^0 = \beta,\quad \gamma_5 = i\, \gamma^0\, \gamma^1\, \gamma^2\, \gamma^3 .
\]
Note that the $\alpha^i \ (i = 1,2, 3)$ anti-commute among themselves and with $\beta ; \gamma_5$ commutes with the $\alpha^i $ and anti-commutes with $\beta$.

With these matrices $K$ in (7) may be presented as 
\begin{equation}
\Psi^\ast\, K \Psi = \Psi^\ast\ ( \alphavec \cdot \bfp + g \beta \ [\varphi^r - i \varphi^i\, \gamma_5 ] )\ \Psi .
\label{rjeq8}
\end{equation}
(Henceforth all vectorial quantities are 2-dimensional.) Here $\bfp$ is the operator $-i {\boldsymbol\nabla}$; we have renamed $\triangle$ as $ g \, \varphi$, where $g$ is a coupling strength and $\varphi$ is a complex scalar field, with real and imaginary parts: $\varphi = \varphi^r + i\, \varphi^i$. Note that when $\varphi$ is decomposed into modulus and phase: $\varphi = \mid\varphi\mid\, e^{i \chi}$, the interaction part of (\ref{rjeq8}) may be presented as $g \mid\varphi\mid\, \Psi^\ast \, \beta\, e^{-i \gamma_5 \chi}\, \Psi$. This makes it clear that this interaction is invariant against a local chiral gauge transformation.
\begin{alignat}{2}
&\varphi \to e^{2 i \omega}\, \varphi \Rightarrow \quad \chi \to \chi+ 2\omega\nonumber\\[1ex]
&\Psi \to e^{i \omega \gamma_5}\, \Psi
\label{rjeq9}
\end{alignat}
[When $\varphi = \varphi_0$ is constant, its constant phase may be removed from K by the above transformation, leaving a conventional Dirac mass term (gap) $\propto \, | \varphi_0 |$. ]

In order that the kinetic portion of (\ref{rjeq8}) be invariant against the local gauge transformation (\ref{rjeq9}), we introduce coupling to a gauge potential  $\Avec $, which transforms as
\begin{equation}
\Avec \to \Avec + {\boldsymbol\nabla} \omega .
\label{rjeq10}
\end{equation}
Thus our final Dirac Hamiltonian density reads
\begin{eqnarray}
\Psi^\ast \, K_A \, \Psi = \Psi^\ast\, \alphavec \cdot [\bfp - \gamma_5\, \Avec] \ \Psi+ g \Psi^\ast \beta \,  [\varphi^r - i\, \gamma_5\, \varphi^i ] \Psi\hspace{.72in}\nonumber\\[1ex]
=\bar{\Psi}_+\, \gammavec \cdot (\bfp - \Avec) \ \Psi_+ + \bar{\Psi}_- \, \gammavec \cdot (\bfp + \Avec) \Psi_-
+ g \, \varphi\, \bar{\Psi}_+ \, \Psi_- + g \varphi^\ast\, \bar{\Psi}_- \, \Psi_+ \, .
\label{rjeq11}
\end{eqnarray}
We have introduced the Dirac adjoint $\bar{\Psi} \equiv \Psi^\ast\, \gamma^0$, and the chiral components $\Psi_\pm \equiv \frac{1}{2} \ (1 \pm \gamma_5) \ \Psi$, whose gauge transformation law is 
\begin{equation}
\Psi_\pm \to e^{\pm i \omega}\ \Psi_\pm , \quad \bar{\Psi}_\pm \to \bar{\Psi}_\pm\, e^{\mp i \omega} .
\label{rjeq12}
\end{equation}

The Bose fields $\varphi$ and $\Avec$ are determined by the familiar NO/LGA equations.
\begin{alignat}{2}
&\Dvec \cdot \Dvec \, \varphi = \varphi V^\prime \, (\varphi^\ast \, \varphi)\nonumber\\[1ex]
& \Dvec \, \varphi \equiv (\nabla - i\, 2 \Avec)\ \varphi
\label{rjeq13}\\[3ex]
& \frac{1}{e^2}\ \varepsilon^{ij}\, \partial_j\, B = j^i_{\mbox{\tiny{BOSE}}}\label{rjeq14}\\[1ex]
&B \equiv \varepsilon^{ij}\, \partial_i\, A^j\nonumber\\[1ex]
&{\bf  j}_{\mbox{\tiny{BOSE}}} = 4 \, I m \, \varphi^\ast\, \Dvec\, \varphi\nonumber
\end{alignat}
Here $e$ is a further coupling constant and $V$ is chosen so that at minimum $V^\prime = 0, \ \varphi^\ast\, \varphi = \varphi^2_0$, which gives rise to the Dirac mass for $\Psi\, \propto |\varphi_0 |$. Furthermore,  vortex solutions to these equations also exist. Their form is
\begin{alignat}{2}
&\varphi_v \, \rparen = \varphi\, (r)\, e^{i n \theta}\, ,\nonumber\\
&A^i_v\, \rparen = -n \, \varepsilon^{ij}\, \frac{r^j}{r^2}\ a\, (r)\, ,
\label{rjeq15}
\end{alignat}
where $n$ is an integer; $\varphi (r)$ vanishes with $r$ as $r^{|n|}$ and tends to $\varphi_0$ at infinity; $a (r)$ vanishes at the origin so that $\Avec_v$ is regular there, and $ a (r)$ tends to 1/2 at large $r$. All this ensures finiteness of the vortex  energy $\int d^2\, r \left(\frac{1}{2e^2}\ B^2 + \mid \Dvec \, \varphi \mid^2 +\, V \, (\varphi^\ast \varphi)\right)$.

Finally, note that our system possesses a global fermion number symmetry, with just the Fermi fields transforming with a constant phase: $\Psi \to e^{i\lambda}\, \Psi$. Consequently the theory possesses a local chiral U(1) symmetry and a global U(1) fermion number symmetry. Because the theory resides in (2+1) dimensions, no chiral anomalies interfere with our chiral gauge symmetry. 

Indeed, in spite of the presence of the $\gamma_5$ matrix, the theory is parity (P) and charge-conjugation (C) invariant and the axial vector current is C-odd. This is because in (2+1) dimension, with 4-component Dirac fields, the relevant transformations read
\begin{alignat}{3}
&P:  \varphi \ (t, x, y) \to \varphi \  (t, -x, y) \nonumber\\[1ex]
&   \quad\ A^{0, y}\ (t, x, y) \to  A^{0, y}\ (t, -x, y) \nonumber\\[1ex]
&\quad\ A^x\  (t, x, y) \to - A^x\,  (t, - x, y)\nonumber\\[1ex]
&\quad\   \Psi\, (t, x, y) \to i\, \gamma^3\, \gamma^1\, \Psi\ (t, -x, y)\label{rjeq16}\\[3ex]
&C: \varphi \to \varphi^\ast\nonumber\\[1ex]
& \quad\ \Avec \to - \Avec \nonumber\\[1ex]
&  \quad\ \Psi_i \to \gamma^1_{ij}\, \bar{\Psi}_j
\label{rjeq17}
\end{alignat}
It follows that time reversal symmetry holds also.
\section{Modified Dirac Equation}
With the additional gauge potential $\Avec$, our Dirac eigenvalue problem differs from HCM. According to  (\ref{rjeq11}) we have
\begin{equation}
(\alphavec \cdot (\bfp -\gamma_5\, \Avec)  + g \beta [\varphi^r -i\, \gamma_5\, \varphi^i])\Psi = E\, \Psi\, .
\label{rjeq18}
\end{equation}
Observe that $ \alpha^3$, which we rename $R$,  anti-commutes with the matrix structure on the left side of (\ref{rjeq18}). Therefore if $\Psi_E$ is an eigenfunction with eigenvalue $E$, $ R\Psi_E$ belongs to eigenvalue $- E$, and zero modes can be chosen as   eigenstates of $R$. This is a consequence of the ``sublattice symmetry" identified by HCM.

Next we show that the gauge interaction in (\ref{rjeq18}) does not affect the zero modes found by HCM without $\Avec$. To this end, we adopt the Coulomb gauge and present $\Avec$ as $A^i = \varepsilon^{ij}\, \partial_j \, A$. Also it is true that $\alpha^i\, \gamma_5 = -i\, \varepsilon^{ij}\, \alpha^j\! R \ (i =1, 2) $. Consequently the kinetic term in (\ref{rjeq18}) can be written as $e^{-A R}\, \alphavec \cdot \bfp \, e^{-A R}$, and (\ref{rjeq18}) becomes 
\begin{equation}
(\alphavec \cdot \bfp + g\, \beta\, [\varphi^r - i \, \gamma_5\, \varphi^i]) \ (e^{-AR}\, \Psi) = E\ (e^{AR} \, \Psi) .
\label{rjeq19}
\end{equation}
Thus $e^{-AR}\, \Psi$ satisfies the HCM equation at $E=0$. Comparison with (\ref{rjeq15}) shows that $A^\prime (r) = - n \, a\, (r) / r$, so that at infinity $A \, (r)$ tends to $-\frac{n}{2}\ l n \, r$, and the zero modes with the gauge interaction differ at large $r$ from the HCM modes by factors $r^{\pm (n/2)}$. This does not affect nomalizability because the zero modes found by HCM are exponentially damped by the interaction with $\varphi_v$. Finally, since the $n= -1$ HCM mode as well as ours has the form,  $ \Psi_0 = \left(
\begin{array}{c}
0\\
v\\
v\\
0
\end{array}
\right)
$, where $v$ is  an exponentially damped function, we see that indeed $\Psi_0$ is an  $R$ eigenstate, with eigenvalue $-1$. Fermion number fractionalization in the gauge theory is now established by the same reasoning as in HCM.

\section{Energy Relations}
The total energy functional for all our fields is 
\begin{eqnarray}
E_{\mbox{\tiny{TOTAL}}} = \int \, d^2\, r \left\{\frac{1}{2\, e^2}\  B^2 + \mid \Dvec \, \varphi\mid^2 + V(\varphi^\ast\, \varphi) \, \right\}
+ \int \, d^2 \, r \, \Psi^\ast \, K_A \, \Psi \, .
\label{rjeq20}
\end{eqnarray}
Varying this with respect to $\Psi^\ast$ produces our Dirac equation (\ref{rjeq18})  at zero eigenvalue. Varying with respect to the Bose fields $\varphi^\ast \ \mbox{and}\ \Avec$ derives (\ref{rjeq13}) and (\ref{rjeq14}), but with a back reaction from the Dirac fields.
\begin{eqnarray}
\Dvec \cdot \Dvec \, \varphi = \varphi\, V^\prime \ (\varphi^\ast \varphi) + \frac{g}{2}\ \bar{\Psi} \ (1+ \gamma_5) \ \Psi\label{rjeq21}\\[1ex]
\frac{1}{e^2}\  \varepsilon^{ij} \, \partial_j\, B = j^i_{\mbox{\tiny{BOSE}}} + \bar{\Psi}\, \gamma^i\, \gamma_5\, \Psi
\label{rjeq22}
\end{eqnarray}
However, with a zero mode that is an eigenstate of R, the back reaction Dirac bilinears vanish. Thus our Dirac zero mode, together with the scalar field/gauge field NO/LGA vortex, is a self consistent solution of the coupled system.

Chiral gauge theories have previously entered physics, but in even-dimensional space-time, where the chiral anomaly influences the structure and physical utility of these models \cite{DJgross}. In the present work, we have a chiral gauge theory in odd-dimensional space-time, whose structure is mathematically very elegant owing to its self-consistent solutions. It remains to be determined whether a microscopic description for graphene can lead to the chiral gauge field that enters our theory.

\section*{Acknowledgement}
We thank C. Chamon and C.-Y. Hou for explaining to us their research on graphene. This work was supported by the Department of Energy under contract numbers DE-FG02-05ER41360 and DE-FG02-91ER40676.

\end{document}